\documentclass[a4paper]{jpconf}
\usepackage{amssymb,latexsym}
\usepackage{amsmath}
\usepackage{amsthm}
\usepackage{graphicx}
\usepackage{tensor}
\usepackage{cite}
\newcommand{\mc}{\mathcal}

\newcommand{\eqr}{\eqref}
\newcommand{\g}{g\indices}
\newcommand{\Q}{\mathcal{Q}}
\newcommand{\A}{\mathcal{A}}
\newcommand{\dr}{\ell}

\theoremstyle{plain}
\theoremstyle{definition}
\newtheorem{assertion}{Assertion}%[section]%[chapter]

\usepackage[colorlinks=true,linkcolor=blue,citecolor=blue,urlcolor=blue]{hyperref}
\begin{document}
%----------------------------------------------------------------------------------------------------------------------
\title{On Quantum Microstates %from the Full Asymptotic Symmetry Group 
in the Near Extremal, Near Horizon Kerr Geometry} 
\author{Ananda Guneratne$^{1}$, Leo Rodriguez$^2$, Sujeev Wickramasekara$^{1,3}$ and Tuna Yildirim$^4$}
\address{$^{1}$Department of Physics and Astronomy, The University of Iowa, Iowa City, IA 52242}
\address{$^2$Department of Physics, Assumption College, Worcester, MA 01609}
\address{$^3$Department of Physics, Grinnell College, Grinnell, IA 50112}
\address{$^4$Department of Physics, Arizona State University, Tempe, AZ 85287}
\ead{\href{mailto:ananda-guneratne@uiowa.edu}{ananda-guneratne@uiowa.edu}}
\ead{\href{mailto:ll.rodriguez@assumption.edu}{ll.rodriguez@assumption.edu}}
\ead{\href{mailto:s-wickram@uiowa.edu}{s-wickram@uiowa.edu}}
\ead{\href{mailto:tuna.yildirim@asu.edu}{tuna.yildirim@asu.edu}}
%$^{a}$\footnote{\href{mailto:rodrigul@grinnell.edu}{rodrigul@grinnell.edu}}, Sujeev Wickramasekara$^{a}$\footnote{\href{mailto:wickrama@grinnell.edu}{wickrama@grinnell.edu}}\\and Tuna Yildirim$^{b}$\footnote{\href{mailto:tuna-yildirim@uiowa.edu}{tuna-yildirim@uiowa.edu}}\vspace{.5cm}\\ \textit{$^{a}$Department of Physics}\\ \textit{Grinnell College}\\\textit{Grinnell, IA 50112}\vspace{.5cm}\\ \textit{$^{b}$Department of Physics and Astronomy}\\ \textit{The University of Iowa}\\\textit{Iowa City, IA 52242}}
%\maketitle
%\date{\today}
\begin{abstract}
We study the thermodynamics of near horizon near extremal Kerr (NHNEK) geometry within the framework of $AdS_2/CFT_1$ correspondence.  We start by shifting the horizon 
of near horizon extremal Kerr (NHEK) geometry by a general finite mass. While this shift does not alter the geometry in that the resulting classical solution is still diffeomorphic to the NHEK  
solution, it does lead to a quantum theory different from that of NHEK.  We obtain this quantum theory by means of a Robinson-Wilczek two-dimensional Kaluza-Klein reduction which enables 
us to introduce a finite regulator on the $AdS_2$ boundary and compute the full asymptotic symmetry group of the two-dimensional quantum conformal field theory on the respective 
$AdS_2$ boundary. The $s$-wave contribution of the energy-momentum-tensor of this conformal field theory, together with the asymptotic symmetries, 
generate a Virasoro algebra with a calculable center, which agrees with the standard Kerr/$CFT$ result, and a non-vanishing lowest Virasoro eigenmode. 
The central charge and lowest eigenmode produce the Bekenstein-Hawking entropy and Hawking temperature for NHNEK. 
%Keywords: Quantum Field Theory in Curved Spacetime, Black Hole Thermodynamics; Black-Hole/CFT Duality; Quantum Gravity.\vspace{.5cm}\\
%PACS numbers: 11.25.Hf, 04.60.-m, 04.70.-s
\end{abstract}
%\begin{center}
%\noindent\line(1,0){150}
%\end{center}
%\newpage
%\thispagestyle{plain}
%----------------------------------------------------------------------------------------------------------------------
%\lhead{\rightmark}
%\chead{HW 10}
%\rhead{\thepage}
%\lfoot{A. Guneratne, L. Rodriguez, S. Wickramasekara, T. Yildirim}
%\cfoot{}
%\rfoot{\thepage}%\noindent We will take the speed of light $c=1$ through out this problem set.
%----------------------------------------------------------------------------------------------------------------------
%\begin{center}
%\rule{\linewidth}{0.5mm}
%\noindent\line(1,0){300}
%\end{center}
%\tableofcontents
%\todototoc
%\listoftodos[List of Questions]
%\begin{center}
%\rule{\linewidth}{0.5mm}
%\noindent\line(1,0){300}
%\end{center}
%\newpage
%------------------------------------------------------------------------------
\section{Introduction}\label{sec:intro}
%------------------------------------------------------------------------------
Black hole thermodynamic quantities \cite{hawk2,hawk3,beken},
\begin{align}
\label{eq:htaen}
\begin{cases}
T_H=\frac{\hbar\kappa}{2\pi}&\text{Hawking Temperature}\\
S_{BH}=\frac{A}{4\hbar G}&\text{Bekenstein-Hawking Entropy}
\end{cases},
\end{align}
have provided a testbed for most current competing theories of quantum gravity. It is widely believed that any viable ultraviolet completion of general relativity should reproduce some variant of \eqr{eq:htaen}, perhaps modulo some real finite parameter that would need to be fixed by experiment. To date, there have been a range of different approaches for arriving at \eqr{eq:htaen}, with string theories and loop quantum gravity being the predominant competitors but no clear consensus on which approach should be preferred over the others.
%--
%\subsection{Black Hole Temperature and Effective Action}\label{sec:bht}
%--

Since Hawking's original analysis of the density of quantum states in terms of Bogolyubov coefficients, effective actions and their associated energy-momentum tensors for semiclassical matter fields have been explored in various settings for arriving at $T_H$\cite{mukwipf,LLRphd}. Of particular interest is the realization by Robinson and Wilczek (RW) that anomalous two-dimensional chiral theories in the near horizon of black holes are rendered unitary by requiring the black hole to radiate at temperature $T_H$ \cite{robwill,isowill,Zampeli:2012tv,srv}. The RW procedure requires a dimensional reduction yielding two-dimensional analogues (RW2DA) for various types of four-dimensional black holes, beyond the basic Schwarzschild case, coupled to two-dimensional matter fields.

%More recently Rodriguez and Yildirim showed by analyzing quantum conformal matter in the background of certain RW2DA black holes that the resulting energy-momentum tensor (EMT) is holomorphic in the horizon limit or at the asymptotic infinity boundary \cite{ry,Button:2010kg}. Furthermore, this resulting holomorphic EMT is dominated by one component equaling the four-dimensional Hawking flux of temperature $T_H$ weighted by a central charge $c$. A plausible interpretation of this result is that in the near horizon regime a four or higher dimensional spacetime metric exhibits a Kaluza-Klein reduction into two-dimensional fields $\g{^{(2)}_\mu_\nu}$, $\mc{A}_\mu$ and $\Phi$. In this interpretation $\mc{A}_\mu$ and $\Phi$ are of gravitational origin, yet mathematically behave as two-dimensional $U(1)$ gauge and conformal scalar fields. Thus a quantum field theoretic study of $\Phi$ in two dimensions with respect to $\g{^{(2)}_\mu_\nu}$ and $\mc{A}_\mu$ will have quantum gravitational implications in the near horizon of four-dimensional black holes. Depending on the asymptotic symmetry group of certain RW2DA, this may suggest a sort of $AdS_2/CFT_1$ relationship for computing four-dimensional black hole temperature.
%\cite{Bardeen:1999px}NHEKciteation
%--
%\subsection{Holographic Black Hole Entropy}\label{sec:bhent}
%--

The $AdS/CFT$ correspondence, i.e., the conjecture that quantum gravity on an anti-de Sitter space is dual to a conformal field theory \cite{Maldacena:1997re},  has spawned a surge, led by Strominger \cite{strom2,kerrcft}, Carlip \cite{Carlip:2011ax,carlip,carlip3,carlip2}, Park \cite{Park:1999tj,Park:2001zn,kkp} and others, in applying $CFT$ techniques to compute Bekenstein-Hawking entropy of various black holes. By far, the most notable example is the Kerr/$CFT$ correspondence and its extensions \cite{kerrcft,Compere:2012jk}, where the general idea is that the asymptotic symmetry group (ASG), preserving certain metric boundary or fall off conditions, is generated by a Virasoro algebra with a calculable central extension:
\begin{align}
\label{eq:vir}
\left[\Q_m,\Q_n\right]=(m-n)\Q_{m+n}+\frac{c}{12}m\left(m^2-1\right)\delta_{m+n,0},
\end{align}
where $m,n\in\mathbb{Z}$. The Bekenstein-Hawking entropy is then obtained from Cardy's formula \cite{cardy2,cardy1} in terms of the central charge $c$ and the 
normalized lowest eigenmode $\Q_0$ (without Casimir shift):
\begin{align}
\label{eq:cf}
S=2\pi\sqrt{\frac{c\cdot\Q_0}{6}}.
\end{align}
Since surface gravity is usually employed in regulating the quantum charges of \eqr{eq:vir}, thus leading to a finite $\Q_0$, there is some difficulty in using \eqr{eq:cf} for the extremal Kerr geometry 
where surface gravity, and therewith Hawking temperature, vanishes. To circumvent this,  a thermal Cardy formula is often employed: 
\begin{align}
\label{eq:tcf}
S=\frac{\pi^2}{3}\left(c_LT_L+c_RT_R\right)
\end{align}
where the subscripts $L$ and $R$ refer to the dual, two chiral CFT theories with central charges $c_L$ and $c_R$ and  
Frolov-Thorne vacuum temperatures $T_L$ and $T_R$ which, in the near extremal 
case, are 
\begin{align}\label{eq:nefttemp}
T_L=\frac{(GM)^2}{2\pi J}\ ~\text{and}~\ T_R=\frac{\sqrt{(GM)^4-(GJ)^2}}{2\pi J}.
\end{align}
It follows that $T_L=\frac{1}{2\pi}$, $T_R=0$ in the extremal case, where $GJ=M^2$. Further, since $c_L=c_R=c=12J$ in the extremal case, 
we can readily obtain  the near horizon extremal Kerr (NHEK) entropy from \eqref{eq:tcf}, which has exactly the same form as the Bekenstein-Hawking entropy:
\begin{align}
\label{eq:tcf2}
S_{BH}=\frac{\pi^2}{3}(c_LT_L+c_RT_R)=\frac{\pi^2}{3}cT_L=2\pi J.
\end{align} 
Because of the equality of $c_L=c_R$, we can also identify a NHEK temperature, 
\begin{equation}
T=T_L+T_R=\frac{1}{2\pi},\label{eq:cfttemp}
\end{equation}
which is in contrast to the vanishing Hawking temperature in this limit. 

Note that, while the use of the thermal Cardy formula was motivated by the failure of regularization of the quantum charges \eqr{eq:vir}, a finite zero mode may be nevertheless 
inferred from \eqref{eq:tcf2} by the identification
\begin{align}\label{eq:fttdef}
\frac{\partial S_{CFT}}{\partial \Q_0}=\frac{\partial S_{BH}}{\partial \Q_0}\equiv\frac{1}{T}\Rightarrow \Q_0=\frac{\pi^2}{6}c T^2.
\end{align}
Turning things around, the temperature $T=\frac{1}{2\pi}$ of \eqref{eq:cfttemp} can be obtained  as a general result  from \eqref{eq:fttdef} when the ASG of the
CFT contains a proper $SL(2,\mathbb{R})$ subgroup. From the definition \eqr{eq:fttdef} we also see that $T$ in general should be unitless (with $\hbar=1$). 
In passing, we note that an interesting candidate for such a general temperature is  the Hawking temperature scaled by the finite time regulator $1/\kappa$. This gives 
$T=\frac{1}{2\pi}$, which also extends smoothly to extremality (a similar identification can be found in \cite{Carlip:2011ax,ChangYoung:2012kd}). 

The desired general temperature result $T=\frac{1}{2\pi}$ motivates us to consider defining a general Kerr entropy that reduces to \eqref{eq:tcf2} at extremality. 
To that end, if we combine the Frolov-Thorne temperatures \eqref{eq:nefttemp} with \eqref{eq:cfttemp} \emph{and} use $c_L=c_R=c=12J$, we can obtain 
the standard area law 
\begin{align}\label{eq:mal}
S_{BH}=2\pi\left(GM^2+\sqrt{G^2M^4-J^2}\right).
\end{align}
The drawback of this  result is that it combines quantities derived separately at extremality and near-extremality. It is also not obvious that the combination of 
temperatures in \eqr{eq:nefttemp}, $T=T_L+T_R$, yields the value $\frac{1}{2\pi}$, except in the extremal limit. However,  since the extremal result $c=12J$ is 
consistent with the general expression $c=\frac{3A}{2\pi G}$, we may 
recast  $c,~T_L,~T_R$ in terms of more general variables to obtain
\begin{align}\label{eq:nexctlr}
c=\frac{3A}{2\pi G},~T_{L}=\frac{4(GM)^2}{A}~\text{and}~T_{R}=\frac{4\sqrt{(GM)^4-(GJ)^2}}{A}.
\end{align}
Substituting \eqref{eq:nexctlr} into \eqr{eq:tcf} yields $S_{BH}=\frac{A}{4G}$ and, assuming they smoothly extend back to non-extremality, 
\begin{align}
T=T_{L}+T_{R}=\frac{4(GM)^2}{A}+\frac{4\sqrt{(GM)^4-(GJ)^2}}{A}=\frac{1}{2\pi}.
\end{align}
Such a procedure would provide a more wholesome calculation of near-extremal Kerr black hole entropy. This is precisely the aim of this note, i.e., to construct a  $CFT$ dual for the 
near horizon, near extremal Kerr (NHNEK) geometry and compute its  entropy by the statistical Cardy formula \eqr{eq:cf}, without mixing results derived separately at extremality and near-extremality. 
This will require the computation of the full ASG, which we will do within an $AdS_2/CFT_1$ correspondence by performing a RW two-dimensional reduction of the NHNEK geometry 
in a specific finite mass gauge, following similar previous constructions \cite{Button:2010kg}. 

\section{Geometry}\label{sec:nensea}
%------------------------------------------------------------------------------
Consider the generic Kerr metric 
\begin{align}
\label{eq:kerrm}
\begin{split}
ds^2_{Kerr}=&-\frac{\Sigma\Delta}{\left(r^2+a^2\right)^2-\Delta a^2\sin^2\theta}dt'^2+\Sigma\left[\frac{dr^2}{\Delta}+d\theta^2\right]\\
&+\frac{\left(\left(r^2+a^2\right)^2-\Delta a^2\sin^2\theta\right)\sin^2\theta}{\Sigma}\left[d\phi'+\frac{2rGMa}{\left(r^2+a^2\right)^2-\Delta a^2\sin^2\theta}dt'\right]^2,
\end{split}
\end{align}
where
\begin{align}
\label{eq:kerrdef}
\begin{split}
\Sigma=&r^2+a^2\cos^2\theta,\\
\Delta=&\left(r-r_+\right)\left(r-r_-\right),\\
r_\pm=&Gm\pm\sqrt{(Gm)^2-a^2},\\
a=&\frac{J}{m}.
\end{split}
\end{align}
The NHEK geometry is a four-dimensional vacuum solution derived from the above Kerr metric 
by introducing the transformations 
\begin{align}
\label{eq:nhekcoor}
r=Gm+\lambda U,~t'=\frac{t}{\lambda},~\phi'=\phi+\frac{t}{2Gm\lambda},
\end{align}
and taking the limit $\lambda\to0$,  known as the extremal near horizon limit: 
\begin{align}
\label{eq:nhek}
ds^2_{NHEK}=\frac{1+\cos^2{\theta}}{2}\left[-\frac{U^2}{\dr^2}dt^2+\frac{\dr^2}{U^2}dU^2+\dr^2d\theta^2\right]+\dr^2\frac{2\sin^2{\theta}}{1+\cos^2{\theta}}\left(d\phi+\frac{U}{\dr^2}dt\right)^2, 
\end{align}
where $\dr^2=2G^2M^2$. This extremal metric  may be tuned to near-extremality via a finite temperature gauge:
\begin{align}
\label{eq:nnhek}
ds^2_{NHNEK}=\frac{1+\cos^2{\theta}}{2}\left[-\frac{U^2-\epsilon^2}{\dr^2}dt^2+\frac{\dr^2}{U^2-\epsilon^2}dU^2+\dr^2d\theta^2\right]+\dr^2\frac{2\sin^2{\theta}}{1+\cos^2{\theta}}\left(d\phi+\frac{U}{\dr^2}dt\right)^2,
\end{align}
where $\epsilon=\frac{1}{2\lambda}\left(r_+-r_-\right)$ is a finite excitation above extremality.  The  line elements \eqref{eq:nhek} and  \eqref{eq:nnhek} are classically diffeomerophic, but they exhibit differing quantum theories.  To extract the details of these differences, we make one more tuning by adding a finite $ADM$ mass parameter. The resulting metric representing this finite mass gauge, 
\begin{align}
\label{eq:knadsnhmkerr}
\begin{split}
ds^2_{NHNEK}=&\frac{1+\cos^2\theta}{2}\left[-\frac{r^2-2rGM+a^2}{r_+^2+a^2}dt^2+\frac{r_+^2+a^2}{r^2-2rGM+a^2}dr^2+\left(r_+^2+a^2\right)d\theta^2\right]+\\
&\frac{2\sin^2\theta}{1+\cos^2\theta}\left(r_+^2+a^2\right)\left[d\phi+\left(\frac{r-2GM}{r_+^2+a^2}\right)dt\right]^2,
\end{split}
\end{align}
is still diffeomorphic to \eqref{eq:nhek} and \eqref{eq:nnhek}, and clearly exhibits global $AdS_2\times S^2$ topology, our reasons for maintaining the same label NHNEK in \eqref{eq:knadsnhmkerr}.
However, it is this form of the NHNEK metric with the mass gauge that will prove useful for the tuning purposes in our $CFT$ construction and the calculation of the full ASG.
%------------------------------------------------------------------------------
\section{Quantum Fields in NHNEK Spacetime}\label{sec:qftnhnek}
%------------------------------------------------------------------------------
Note that the NHNEK metric \eqref{eq:nhek} is of the general form 
\begin{align}
\label{eq:knadsnhm}
ds^2=K_1\left(\theta\right)\g{^{(2)}_\mu_\nu}dx^\mu dx^\nu+K_2\left(\theta\right)e^{-2\varphi}d\theta^2+K_3\left(\theta\right)e^{-2\varphi}\left[d\phi+\mathcal{A}\right]^2,
\end{align} 
where $\g{^{(2)}_\mu_\nu}$ is the reduced two-dimensional metric $(\mu,\nu=0,1)$, $\varphi$ is a free scalar field and $\mc{A}$ is a $U(1)$ gauge field, collectively often called 
Kaluza-Klein fields. The two-dimensional field splitting of \eqref{eq:knadsnhm} provides a robust platform for constructing $CFT$ duals in the near extremal case by way of the ASG of the Kaluza-Klein fields. We will now exploit the structure of \eqref{eq:knadsnhm} to examine the near horizon matter theory of \eqref{eq:knadsnhmkerr} via the RW dimensional reduction procedure. Our goal here is to apply our previous techniques from \cite{ry,Button:2010kg,Button:2013rfa} to study the resulting thermodynamics of the NHNEK within an $AdS_2/CFT_1$ formalism. 
%--
\subsection{RW Dimensional Reduction}\label{sec:RWdr}
%--
Our initial ansatz leading to the NHNEK solution involved a specific decomposition of our four-dimensional spacetime into a two-dimensional black hole and matter fields. 
It is necessary to check  that these fields are the correct RW2DA useful in a holographic study of the quantum spacetime in the near horizon regime. To that end, 
let us consider a single free scalar field $\varphi$ in the background of \eqr{eq:knadsnhmkerr} with action:
\begin{align}
\label{eq:freescalar4}
\begin{split}
S_{free}=&\frac12\int d^4x\sqrt{-g}g^{\mu\nu}\partial_{\mu}\varphi\partial_\nu\varphi\\
=&-\frac12\int d^4x\,\varphi\left[\partial_\mu\left(\sqrt{-g}g^{\mu\nu}\partial_\nu\right)\right]\varphi\\
=&-\frac12\int d^4x\,\varphi\left[\partial_t\left( -\sin{\theta}\left(a^2+r_+^2\right)\frac{r_+^2+a^2}{r^2-2rGM+a^2}\partial_t\right)+\right.\\
&\ \partial_r\left( \sin{\theta}\left(a^2+r_+^2\right)\frac{r^2-2rGM+a^2}{r_+^2+a^2}\partial_r\right)+\partial_\theta\left( \sin{\theta}\partial_\theta\right)+\\
&\ \ \partial_\phi\left( \left\{-\sin{\theta}\left(a^2+r_+^2\right)\left(\frac{r-2GM}{r_+^2+a^2}\right)^2\frac{r_+^2+a^2}{r^2-2rGM+a^2}+\frac{(\cos{(2 \theta )}+3)^2}{16\sin\theta}\right\}\partial_\phi\right)+\\
&\ \ \ \left.2\partial_t\left( \sin{\theta}\left(a^2+r_+^2\right)\frac{r-2GM}{r_+^2+a^2}\frac{r_+^2+a^2}{r^2-2rGM+a^2}\partial_\phi\right)\right]\varphi.
\end{split}
\end{align}
The above functional is reduced to a two-dimensional theory by expanding the four-dimensional scalar field in terms of spherical harmonics
\begin{align}
\label{eq:sphdecom}
\varphi(t,r,\theta,\phi)=\sum_{lm}\varphi_{lm}(r,t)Y\indices{_l^m}(\theta,\phi),
\end{align}
where $\varphi_{lm}$ has the form of a complex interacting two-dimensional scalar field. Integrating out angular degrees of freedom, transforming to tortoise coordinates $dr^*=f(r)dr$ and considering the region very close to $r_+$, we find that the two-dimensional action is much reduced. This is due to the fact that all interaction, mixing and potential terms ($\sim l(l+1)\ldots$) are weighted by a factor of $f(r(r*))\sim e^{2\kappa r^{*}}$, which vanishes exponentially fast as $r\to r_+$. This leaves us with an infinite collection of massless charged scalar fields in the very near horizon region, with $U(1)$ gauge charge equal to the azimuthal quantum number $e=m$ and remnant functional:
\begin{align}
\label{eq:nhapw}
S=-\frac{r_+^2+a^2}{2}\int d^2x\;\varphi^{*}_{lm}\left[-\frac{1}{f(r)}\left(\partial_t-im\A_t\right)^2+\partial_rf(r)\partial_r\right]\varphi_{lm}.
\end{align}
Thus, we arrive at the RW2DA for the NHNEK solution given by:
\begin{align}
\label{eq:2drwamet}
\g{^{(2)}_\mu_\nu}=\left(\begin{array}{cc}-f(r) & 0 \\0 & \frac{1}{f(r)}\end{array}\right)
\end{align}
and $U(1)$ gauge field
\begin{align}
\label{eq:rw2dgf}
\A=\A_tdt.
\end{align}

Given the initial ansatz \eqr{eq:knadsnhm}, it is not surprising that the only relevant physical fields in the region $r\sim r_+$ are the above RW2DAs, which reinforces 
 the holographic statement that we may learn much about the quantum nature of spacetime in the near horizon regime from the semiclassical analysis of $\g{^{(2)}_\mu_\nu}$, $\A$ and $\varphi_{lm}$.
%--
\subsection{Effective Gravitational Action and Asymptotic Symmetries}\label{sec:asefa}
%--
We would like to interpret \eqr{eq:nhapw} as a useful action for gravity in the near horizon of the classical four-dimensional spacetime. This can be done by considering only the $s$-wave contribution and making a field redefinition rendering the scalar field unitless \cite{Button:2013rfa}. The $s$-wave approximation is sensible in this scenario since we will interpret $\varphi_{lm}$ as a component of the gravitational field and hence it should be real and unitless. Most of the interesting gravitational dynamics seem to be contained in this region or approximation \cite{strom1}. We also note that in \cite{Button:2010kg} it was shown that $\varphi_{lm}$ dies exponentially fast in time by analyzing the asymptotic behavior of its field equation. However, we find the statement relating $\varphi_{lm}$ to a real gravitational field component, a stronger justification to neglect higher order terms in $l$ and $m$. These arguments motivate the field redefinition
\begin{align}\label{eq:sfrd}
\varphi_{00}=\sqrt{\frac{6}{G}}\psi,
\end{align} 
where $\psi$ is now unitless and the $\sqrt{6}$ is chosen to recover the Einstein coupling $\frac{1}{16\pi G}$ in the effective action  \eqr{eq:nhapw} within the $s$-wave approximation.
Using \eqr{eq:sfrd} in \eqr{eq:nhapw} gives
\begin{align}\label{eq:tdtrw2}
S^{(2)}[\psi,g]=\frac{3(r_+^2+a^2)}{G}\int d^2x\sqrt{-g^{(2)}}\psi\left[D_{\mu}\left(\sqrt{-g^{(2)}}\g{_{(2)}^{\mu\nu}}D_{\nu}\right)\right]\psi,
\end{align}
where $D_\mu$ is the gauge covariant derivative. In addition to re-dressing the scalar field, our choice of field redefinition has also rendered the effective coupling unitless, hinting towards a finite quantum theory. The effective action of this quantum theory, which may be extracted via zeta-function regularization of the functional determinant in \eqr{eq:tdtrw2}, is given by the sum of two functionals \cite{isowill,Leutwyler:1984nd}:
\begin{align}\label{eq:nhpcft}
\Gamma=&\Gamma_{grav}+\Gamma_{U(1)},
\end{align}
where
\begin{align}
\begin{split}
\Gamma_{grav}=&\frac{(r_+^2+a^2)}{16\pi G}\int d^2x\sqrt{-g^{(2)}}R^{(2)}\frac{1}{\square_{g^{(2)}}}R^{(2)}\\
\Gamma_{U(1)}=&\frac{3 e^2 (r_+^2+a^2)}{\pi G}\int \mc{F}\frac{1}{\square_{g^{(2)}}}\mc{F},
\end{split}
\end{align}
and $R^{(2)}$ is the Ricci scalar curvature obtained from $\g{^{(2)}_{\mu\nu}}$, and $\mc{F}=d\mc{A}$ is the $U(1)$ invariant curvature two form. Next,  let us introduce the auxiliary scalars $\Phi$ and $B$ satisfying:
\begin{align}\label{eq:afeqm}
\square_{g^{(2)}} \Phi=R^{(2)}~\mbox{and}~\square_{g^{(2)}} B=\epsilon^{\mu\nu}\partial_\mu \A_\nu,
\end{align}
which transform the functional \eqr{eq:nhpcft} into a Liouville $CFT$ of the form:
\begin{align}\label{eq:nhlcft}
\begin{split}
S_{NHCFT}=&\frac{(r_+^2+a^2)}{16\pi G}\int d^2x\sqrt{-g^{(2)}}\left\{-\Phi\square_{g^{(2)}}\Phi+2\Phi R^{(2)}\right\}\\
&+\frac{3 e^2 (r_+^2+a^2)}{\pi G}\int d^2x\sqrt{-g^{(2)}}\left\{-B\square_{g^{(2)}}B\right.\\
&+\left.2B \left(\frac{\epsilon^{\mu\nu}}{\sqrt{-g^{(2)}}}\right)\partial_\mu A_\nu\right\}
\end{split}
\end{align}

Now, we turn our attention to computing the ASG of \eqr{eq:tdtrw2}. The behavior of the RW2DA fields at large $r$ is defined by
\begin{align}\label{eq:as2dwa}
\g{^{(0)}_\mu_\nu}=&
\left(
\begin{array}{cc}
-\frac{r^2}{\dr^2}+\frac{2 r G M}{\dr^2}-\frac{a^2}{\dr^2}+\mathcal{O}\left(\left(\frac{1}{r}\right)^3\right)& 0 \\
 0 & \frac{\dr^2}{r^2}+\mathcal{O}\left(\left(\frac{1}{r}\right)^3\right) 
\end{array}
\right),\\
\label{eq:asgf}
\mathcal{A}\indices{^{(0)}_t}=&\frac{r}{\dr^2}-\frac{2 G M}{\dr^2}+\mathcal{O}\left(\left(\frac{1}{r}\right)^3\right),
\end{align}
which yield an asymptotically $AdS_2$ configuration with Ricci Scalar, $R=-\frac{2}{l^2}+O\left(\left(\frac{1}{r}\right)^1\right)$, where $\dr^2=r_+^2+a^2$. In addition, 
we impose the following metric and gauge field fall-off conditions:
\begin{align}\label{eq:mbc}
\delta g_{\mu\nu}=
\left(
\begin{array}{cc}
    \mathcal{O}\left(\left(\frac{1}{r}\right)^3\right)&
   \mathcal{O}\left(\left(\frac{1}{r}\right)^0\right) \\
 \mathcal{O}\left(\left(\frac{1}{r}\right)^0\right) &
\mathcal{O}\left(r\right) 
\end{array}
\right)~\mbox{and}~\delta \mathcal{A}=\mathcal{O}\left(\left(\frac{1}{r}\right)^0\right),
\end{align}
which imply the following set of asymptotic metric preserving diffeomorphisms
\begin{align}\label{eq:dpr}
\xi_n=\xi_1(r)\frac{e^{i n \kappa\left(t\pm r^*\right)}}{\kappa}\partial_t+\xi_2(r)\frac{e^{i n \kappa\left(t\pm r^*\right)}}{\kappa}\partial_r,
\end{align}
where $r^*$ is the tortoise coordinate, 
\begin{align}
\xi_1=C r e^{i n \kappa  r^*},~\xi_2=\frac{i r C (r-G M)}{\kappa  n \left(r^2-2 r G M+a^2\right)},
\end{align}
$C$ is an arbitrary normalization constant and $\kappa$ is the surface gravity of the NHNEK black hole. Under diffeomorphisms \eqref{eq:dpr}, the gauge field transforms as:
\begin{align}
\delta_\xi \mathcal{A}_\mu=\left(\mathcal{O}\left(\left(\frac{1}{r}\right)^0\right),\mathcal{O}\left(\left(\frac{1}{r}\right)^1\right)\right)
\end{align}
and thus $\delta_{\xi}$ may be elevated to a total symmetry of the action, i.e.,
\begin{align}
\delta_\xi\rightarrow\delta_{\xi+\Lambda},
\end{align}
in accordance with \eqr{eq:mbc}. Switching to light cone coordinates $x^\pm=t\pm r^*$ (where large $r$ behavior will be synonymous with large $x^+$ behavior),
we see that the set $\xi_n^\pm$ is well-behaved on the $r\rightarrow\infty$ boundary and form a centerless Witt or $Diff(S^1)$ subalgebra:
\begin{align}
i\left\{\mathbf{\xi}^\pm_m,\mathbf{\xi}^\pm_n\right\}=(m-n)\mathbf{\xi}^\pm_{m+n}.
\end{align}
%--
\subsection{Energy-Momentum and the full ASG}\label{sec:thermo}
%--
We define the energy-momentum tensor and $U(1)$ current of \eqr{eq:nhlcft} in the usual way:
\begin{align}
\label{eq:emt}
\begin{split}
\left\langle T_{\mu\nu}\right\rangle=&\frac{2}{\sqrt{-g^{(2)}}}\frac{\delta S_{NHCFT}}{\delta g\indices{^{(2)}^\mu^\nu}}\\
=&\frac{r_+^2+a^2}{8\pi G}\left\{\partial_\mu\Phi\partial_\nu\Phi-2\nabla_\mu\partial_\nu\Phi+g\indices{^{(2)}_\mu_\nu}\left[2R^{(2)}-\frac12\nabla_\alpha\Phi\nabla^\alpha\Phi\right]\right\}\\
&+\frac{6 e^2 (r_+^2+a^2)}{\pi G}\left\{\partial_\mu B\partial_\nu B-\frac12\g{_\mu_\nu}\partial_\alpha B\partial^\alpha B\right\}~\mbox{and}\\
\left\langle J^{\mu}\right\rangle=&\frac{1}{\sqrt{-g^{(2)}}}\frac{\delta S_{NHCFT}}{\delta \mathcal{A}_\mu}=\frac{6 e^2 (r_+^2+a^2)}{\pi G}\frac{1}{\sqrt{-g^{(2)}}}\epsilon^{\mu\nu}\partial_\nu B.
\end{split}
\end{align}
Next, solving the equations of motions for the auxiliary fields,
\begin{align}
\label{eq:eqmp}
\begin{split}
\square_{g^{(2)}}\Phi=&R^{(2)}\\
\square_{g^{(2)}}B=&\epsilon^{\mu\nu}\partial_\mu \mathcal{A}_\nu
\end{split}
\end{align}
using the metric \eqr{eq:2drwamet} and gauge field \eqr{eq:rw2dgf} and employing the modified Unruh vacuum boundary conditions \cite{unruh}
\begin{align}
\label{eq:ubc}
\begin{cases}
\left\langle T_{++}\right\rangle=\left\langle J_{+}\right\rangle=0&r\rightarrow\infty,~\dr\rightarrow\infty\\
\left\langle T_{--}\right\rangle=\left\langle J_{-}\right\rangle=0&r\rightarrow r_+
\end{cases},
\end{align}
we determine all relevant integration constants of \eqr{eq:emt} and \eqr{eq:eqmp}. For large $r$ and  to $\mathcal{O}(\frac{1}{\dr})^2$, which we will denote as the single limit $r\to\infty$ in the remainder of this section, the resulting energy-momentum tensor is dominated by one holomorphic component, $\left\langle T_{--}\right\rangle$. We are interested in the response of the energy-momentum tensor and the $U(1)$ current to total symmetry $\delta_{\xi^-_n+\Lambda}$, which may be obtained by the use of the boundary fields \eqr{eq:as2dwa} and \eqr{eq:asgf}:
\begin{align}
\begin{cases}
\delta_{\xi^-_n+\Lambda}\left\langle T_{--}\right\rangle=\xi^-_n\left\langle T_{--}\right\rangle'+2\left\langle T_{--}\right\rangle\left(\xi^-_n\right)'+\frac{r_+^2+a^2}{4\pi G}\left(\xi^-_n\right)'''+\mathcal{O}\left(\left(\frac{1}{r}\right)^3\right)\\
\delta_{\xi^-_n+\Lambda}\left\langle J_{-}\right\rangle=\mathcal{O}\left(\left(\frac{1}{r}\right)^3\right).
\end{cases}
\end{align}
From this, we see that $\left\langle T_{--}\right\rangle$ transforms asymptotically as the energy-momentum tensor of a one dimensional $CFT$ with center:
\begin{align}\label{eq:center}
\frac{c}{24\pi}=\frac{r_+^2+a^2}{4\pi G}\Rightarrow c=\frac{3A}{2\pi G}.
\end{align}
We should also note that the above central charge is in congruence with the 2-dimensional conformal/trace anomaly \cite{cft}:
\begin{align}
\label{eq:tra}
\left\langle T\indices{_\mu^\mu}\right\rangle=-\frac{c}{24\pi}R^{(2)}.
\end{align}
%--
\subsection{Virasoro Algebra}\label{sec:vassal}
%--
Next, we define the quantum generators from the conserved charges:
\begin{align}
\label{eq:ccppb}
\Q_n=\lim_{r\rightarrow\infty}\int dx^-\left\langle T_{--}\right\rangle\mathbf{\xi}^-_n,
\end{align}
The algebraic structure of these generators can be found by calculating  response of \eqref{eq:ccppb} to a total symmetry, 
while compactifying the $x^-$ coordinate to a circle from $0\to2\pi/\kappa$:
\begin{align}
\label{eq:ca}
\delta_{\xi^-_m+\Lambda}\Q_n=\left[\Q_m,\Q_n\right]=(m-n)\Q_n+\frac{c}{12}m\left(m^2-1\right)\delta_{m+n,0},
\end{align}
Hence,  the quantum symmetry generators form a centrally extended Virasoro algebra with regulated/normalized zero-mode $\Q_0=\frac{A}{16\pi G}$.
%--
\subsection{$AdS_2/CFT_1$ and Entropy of NHNEK}\label{sec:ent}
%--
The main conclusion of the foregoing discussion is that,  by employing the finite mass gauge \eqref{eq:knadsnhmkerr}, it is possible to show that the near-extremal Kerr throat is
 holographically dual to a $CFT$ with center
\begin{align}\label{eq:cre}
c&=\frac{3A}{2\pi G}
\end{align}
and lowest Virasoro eigenmode 
\begin{align}
\Q_0&=\frac{A}{16\pi G}.
\end{align}
We are now free to use the above results in the statistical Cardy Formula \eqr{eq:cf},
\begin{align}
S=2\pi\sqrt{\frac{c\Q_0}{6}}=\frac{A}{4G}=2\pi\left(GM^2+\sqrt{G^2M^4-J^2}\right),
\end{align}
which is in agreement with the area law \eqr{eq:mal}. However, here we have  derived it without mixing results computed separately at near-extremality and extremality. 
In addition, $c$ and $\Q_0$ extend smoothly to extremality in the limit as $a\to GM$, giving 
\begin{align}
\lim_{a\to GM}c=12J~\text{and}~\lim_{a\to GM}\Q_0=J/2. 
\end{align}
The limiting value $c$ here is identical to the left central charge obtained in the Kerr/$CFT$ correspondence \cite{kerrcft} and, together with the limit of $\Q_0$, it reproduces the extremal
 Kerr entropy by way of the statistical Cardy formula \eqref{eq:cf}. In addition, our derived zero-mode in \eqr{eq:cre} is in accordance with the following assertion \cite{Button:2010kg}:
  \begin{assertion}\label{ass:zerom}
\emph{The lowest Virasoro eigenmode of a quantum $CFT$ is proportional to the irreducible mass of its dual black hole,}
\begin{align}\label{eq:irrmq0}
\mathcal{Q}_0=GM_{irr}^2.
\end{align}
\end{assertion}
\noindent Note that the irreducible mass is the final $ADM$ mass state of a Kerr black hole after it has completed its Penrose process. 
This assertion may have broad generality with a large avenue of application, though we do not have a rigorous proof at this time. 
%--
\subsection{Near Extremal Black Hole Temperature}\label{sec:temp}
%--
To extract the NHNEK temperature, we will focus on the gravitational part of \eqr{eq:nhlcft}, i.e.,
\begin{align}\label{eq:nhlcftgrav}
S_{grav}=&\frac{(r_+^2+a^2)}{16\pi G}\int d^2x\sqrt{-g^{(2)}}\left\{-\Phi\square_{g^{(2)}}\Phi+2\Phi R^{(2)}\right\}.
\end{align}
%\begin{align}
%\label{eq:louiacttem}
%S_{Liouville}=\frac{1}{96\pi}\int d^2x\sqrt{-g^{(2)}}\left\{-\Phi\square_{g^{(2)}}\Phi+2\Phi R^{(2)}\right\}.
%\end{align}
The energy-momentum is given by:
\begin{align}
\label{eq:emtgra}
\begin{split}
\left\langle T_{\mu\nu}\right\rangle=&\frac{2}{\sqrt{-g^{(2)}}}\frac{\delta S_{NHCFT}}{\delta g\indices{^{(2)}^\mu^\nu}}\\
=&\frac{r_+^2+a^2}{8\pi G}\left\{\partial_\mu\Phi\partial_\nu\Phi-2\nabla_\mu\partial_\nu\Phi+g\indices{^{(2)}_\mu_\nu}\left[2R^{(2)}-\frac12\nabla_\alpha\Phi\nabla^\alpha\Phi\right]\right\},
\end{split}
\end{align}
which, as before, may be brought to the form  \eqr{eq:ubc} by using \eqr{eq:emt}. However,  on the horizon limit $r\to r_+$, we are left with just one holomorphic 
component, 
\begin{align}
\label{eq:hhf}
\left\langle T_{++}\right\rangle=-\frac{r_+^2+a^2}{32 \pi G }f'\left(r^+\right)^2.
\end{align}
This is precisely the Hawking flux ($HF$, radiation flux $\sim T\indices{_r^t}$) of the NHNEK metric, weighted by the central charge \eqr{eq:center}, 
\begin{align}
\label{eq:htfhf}
\left\langle T_{++}\right\rangle=cHF=-c\frac{\pi}{12}\left(T_H\right)^2
\end{align}
with Hawking temperature\cite{Jinwu,caldarelli:1999xj}
\begin{align}
T_H=\frac{f'\left(r^+\right)}{4 \pi }.
\end{align}
This is an interesting result, for it suggests that the $AdS_2/CFT_1$ correspondence constructed here contains information about both black hole entropy and black hole temperature. Though the $\left\langle T_{++}\right\rangle$ component in the horizon limit is not precisely the Hawking flux of the four-dimensional parent black hole, given prior knowledge of the central extension, it is possible to read off or extract the relevant information from the correspondence.
%--------------------------------------------------------------------------
\section{Concluding Remarks}\label{sec:concom}
%------------------------------------------------------------------------------
We have analyzed quantum near-extremal Kerr black hole properties in the near horizon regime by way of $AdS_2/CFT_1$ correspondence, as outlined in Table~\ref{tb:adscftc}. This extends our previous work \cite{ry,Button:2010kg} to a new spacetime geometry. The main results of this work includes the derivation of the central charge $c=\frac{3A}{2\pi G}$ from a Lagrangian analysis of conserved currents of two near horizon theories and the correct quadratic two-form transformation law on the holographic renormalized boundary.
\begin{table}[htbp]
\caption{\label{tb:adscftc}Black-Hole/Near-Horizon-$CFT$ Duality}
\begin{center}
\begin{tabular}{ll}
\br
%\multicolumn{5}{|c|}{Black Hole Comparison}\\\hline\hline
$CFT$&Black Hole \\ \mr
Conformal Group&Asymptotic Symmetry Group \\ %\hline
center&$\frac{3A}{2\pi G}$ \\ %\hline
Hamiltonian eigenvalue&$GM^2_{irr}$ \\ %\hline
Regulator&$\kappa_{NHNEK}$ \\ \br
\end{tabular}
\end{center}
\end{table}

It is conceivable that other $AdS_2\times S^2$ gauges, exhibiting the field splitting of \eqr{eq:knadsnhm}, exist with physical relevance and connections to other classical near-extremal solutions. Many analogues to the NHEK for charged rotating black holes with negative and positive cosmological constants have been shown to exist (see \cite{Compere:2012jk} for a comprehensive review). 
They suggest that there may exist similar such analogues to the NHNEK solution, to which the line of reasoning of this note  may be applicable. In particular, Assertion~\ref{ass:zerom} may be a useful tool in the asymptotic symmetry analysis of other extremal black holes. They have vanishing surface gravity, a complication that is generally handled by the use of the thermal Cardy formula \eqr{eq:tcf}. However,  extremal black holes in general do have well defined horizons, which lead to a finite non-zero irreducible mass. This fact enables us to invoke Assertion~\ref{ass:zerom} to 
 implement the standard statistical Cardy formula \eqr{eq:cf} and obtain the Bekenstein-Hawking entropy for a wide class of extremal black holes.
%------------------------------------------------------------------------------
\ack
We thank Vincent Rodgers, Jacob Willig-Onwuachi, John Baker and Shanshan Rodriguez for enlightening discussions. SW thanks the organizers of QuantumFest 2015 for their 
kind invitation and extraordinary hospitality during the conference. This work was supported in part by the HHMI  Science Education 
Award 52006298 and the Grinnell College  CSFS and MAP programs.

This work was completed just before the passing of Sujeev Wickramasekara on December 28th, 2015. Sujeev was our friend, mentor, teacher and collaborator. A true visionary and inspiration in his exemplification of a teacher-scholar and human being! He more than inspired us to pursue greatness in all aspects of our life goals. His teachings and mentorship have and will continue to influence and outline our lives. He is dearly missed and remembered always$\ldots$

%\headheight=16pt
%\rhead{Rodriguez, \textbf{\textit{D\&G}}}
%\lhead{\the\month/\the\day/\the\year }
%\lhead{L. Rodriguez}
%\chead{Notes on $M\cap\Phi$}
%\setlength{\unitlength}{1mm}
%\begin{fmffile}{fmftempl}
%------------------------------------------------------------------------
%\tableofcontents 
%\begin{center}
%\noindent\line(1,0){150}
%\end{center}

%----------------------------------------
\vspace{.5cm}
\begin{center}
\noindent\line(1,0){150}
\end{center}
%---------------------------------
\bibliographystyle{iopart-num}
\bibliography{cftgr}
%\nocite{*}
%---------------------------------

\end{document}